\documentstyle[prl,aps,]{revtex}
\draft
\input epsf
\begin{document}

\title{Guiding neutral atoms around curves \\
with lithographically patterned current-carrying wires}

\author{Dirk M\"{u}ller,
\renewcommand{\thefootnote}{\fnsymbol{footnote}}Dana Z. Anderson, Randal J. Grow, Peter D. D. Schwindt, Eric A.
Cornell\footnote[1]{Staff member, Quantum Physics Division,
National Institute of Standards and Technology, Boulder, CO 80309}
\renewcommand{\thefootnote}{\arabic{footnote}} }

\address{Department of Physics and JILA, University of Colorado and National
Institute of Standards and Technology,\\ Boulder, Colorado
80309-0440}

\date{\today}

\maketitle

\begin{abstract}
Laser-cooled neutral atoms from a low-velocity atomic source are
guided via a magnetic field generated between two parallel wires
on a glass substrate.  The atoms bend around three curves, each
with a 15-cm radius of curvature, while traveling along a
10-cm-long track. A maximum flux of $2\cdot10^{6} \rm{atoms/sec}$
is achieved with a current density of $3\cdot10^{4} \rm{A/cm^{2}}$
in the $100\times100$-$\rm{\mu m}$-cross-section wires.  The
kinetic energy of the guided atoms in one transverse dimension is
measured to be 42 $\rm{\mu K}$.
\end{abstract}

\par

Just as optical waveguides play a central role in many aspects of
modern optics, from communications to integrated optics, atom
waveguides are likely to be an enabling technology for future
atom-optics-based science.  In particular, well-characterized atom
waveguides may make possible inertial and rotation measurements of
exquisite sensitivity via large-enclosed-area atom interferometers
\cite{atom interferometry}.  One of the first guides for atoms was
based on optical forces, where hollow glass fibers guide light,
and the light in turn guides atoms \cite{Renn 1,Ito 1}.

Atom guiding using magnetic forces from current-carrying wires has
been demonstrated more recently \cite{Hinds,Denschlag}.  From the
point-of-view of using atom-guides to pursue precision metrology
goals \cite{atom optics} much benefit can be derived from
patterning the waveguides on a rigid substrate. First,
beamsplitters can be precisely and reproducibly fabricated.
Second, the enclosed area of an interferometer can be precisely
controlled.  Third, the use of well-established lithographic
techniques means that progress on individual optical elements (a
beamsplitter, or a monochromator, for instance) can be rapidly
extended to multi-component experiments.  Mirrors based on
micro-patterned wires have already been introduced \cite{magnetic
mirror}. We report magnetic guiding by a pair of parallel wires
produced on a glass substrate by photolithography and subsequent
electro-plating. Intricate two-dimensional guiding structures are
easily produced by this manufacturing technique; in the present
case, it allows us to demonstrate guiding around curves in a 10 cm
long guide \cite{magnetic guide}.

We guide weak-field-seeking atoms along a one-dimensional
magnetic-field minimum. Our magnetic field is produced by two
parallel wires with equal currents in the same direction. The
track consists of two $100\times100$ $\rm{\mu m}$ wires spaced 200
microns from center to center, providing a 100 $\rm{\mu m}$ space
between the wires. The resulting magnetic field is zero at the
center between the wires and increases linearly outward. A small
longitudinal field is applied to prevent the field magnitude from
vanishing at the track center. The maximum transverse guiding
potential increases linearly with applied current. The transverse
magnetic-field gradient around the center is proportional to the
track current and inversely proportional to the spacing between
the wires. The wire spacing, applied wire current, and the
longitudinal velocity of the guided atoms determine the minimum
radius of curvature around which the atoms can be bent.

Our experimental apparatus consists of two chambers connected by a
2-inch-diameter steel tube that holds the substrate of the wire
guide as shown in figure 1. The chambers are evacuated by separate
ion pumps that maintain the pressure in each chamber typically at
$10^{-9}$ Torr. The source chamber provides a beam of laser-cooled
atoms, and the detection chamber houses a hot wire and channeltron
electron multiplier to measure the atom flux.

A modified magneto-optical trap (MOT) in the source chamber
produces the beam of laser-cooled atoms \cite{MOT}. A diode laser
in a master-oscillator power-amplifier configuration (MOPA)
\cite{MOPA} provides 350 mW of single-frequency light tuned near
the $5S_{1/2}(F=2)\rightarrow 5P_{3/2}(F'=3)$ transition in
rubidium for trapping and cooling in the MOT.  This light is
divided into three beams, which are directed into the chamber
along orthogonal axes, and retro-reflected to supply cooling along
all directions. A 30-mW external-cavity diode laser \cite{diode
laser} supplies light tuned to the $5S_{1/2}(F=1)\rightarrow
5P_{3/2}(F'=2)$ transition to repump atoms that fall into the
$F=1$ ground state back into the cycling transition.  A
500-$\rm{\mu m}$ hole is drilled in the center of one of the
retro-reflecting mirrors, and this mirror is placed inside the
vacuum chamber.  Thus, one of the six confining laser beams has a
dark region in the center of its cross-section. The
radiation-pressure imbalance for atoms in the MOT that enter into
the shadow of the hole accelerates those atoms toward and then
through the hole in the mirror. The resulting atomic beam is
referred to as a low-velocity intense source (LVIS) \cite{LVIS}.
Our observations show that $90\%$ of the LVIS flux atoms are
optically pumped into the $F=1$ ground state by the MOT light.  We
observe that roughly $50\%$ are in the $m_F=0$ state and the rest
of the atoms are roughly equally divided between the two
$m_F=\pm1$ sublevels. Therefore, only $25\%$ of LVIS atoms are in
the correct state to be guided. Typically, we measure an overall
LVIS flux of $\sim 5 \cdot 10^8$ atoms/sec and a beam brightness
of $\sim 5 \cdot 10^{12}\rm{atoms/sr \cdot sec}$. We estimate the
transverse-velocity distribution entering our guide to be about
$v_t=5.0 \pm 2.0$ cm/sec\cite{LVIS}. A time-of-flight measurement
found the longitudinal velocity of LVIS to be $v_l = 10.1 \pm 2.0$
m/sec .

Our tracks were manufactured by Metrigraphics \cite{Metrigraphics}
using photolithography and electro-plating techniques.  A layer of
photoresist is applied on top of a 3-$\rm{\mu m}$-thick layer of
copper deposited onto a $10\times10$ cm glass substrate (Fig. 2).
 This photoresist is then exposed through a mask and removed where
the tracks will be grown.  Using an electro-plating technique, the
tracks are grown through the gaps in the photoresist to a height
of 100 microns.  The excess photoresist and 3-$\rm{\mu m}$ copper
layer are removed, leaving behind a track structure. The final
track structure extends $10$ cm. We solder current-feeding wires
to connection pads on the glass substrate.

The tracks are aligned with the mirror hole before the vacuum
chambers are assembled. We define the track axis as the line
joining the beginning and the end of the track (Fig. 2). The
beginning of the track sits 1 mm behind the mirror hole and its
axis is aligned parallel to the mirror axis (Fig. 1). The magnetic
guide starts with a 1.2 cm straight region followed by a
1.9-cm-long curve to the right with a 15 cm radius of curvature
(Fig. 2). The curvature is then reversed for a 3.8-cm-long left
curve. This region is followed by a 1.9-cm-long right curve,
completing the three alternating curves and leading back to the
track axis. The three curves lead atoms around a bend, diverting
their trajectory by 2 mm transverse to the track axis. After the
bend, the guide confines atoms for another 1.2 cm to a straight
trajectory before they exit the magnetic guide and travel 7 cm
through free space to the detector hot wire.  We place a glass
barrier halfway along the track axis (Fig. 2) to block out direct
LVIS flux. Guided atoms are led around this barrier by the
magnetic guide and can be detected downstream.  For our guiding
experiment we run 35-msec-long current pulses of up to 4.5 A
through the two wires. We choose short current pulses to prevent
the substrate from overheating, allowing us to run larger guiding
currents than continuous currents would allow.

After exiting the guide, atoms are ionized by the hot wire and the
subsequent ions are then detected by the channeltron. The
70-$\rm{\mu m}$-diameter hot wire placed $\sim7$ cm from the
output of the magnetic guide intercepts a small fraction of the
diverging atomic beam.  We determine the total flux from our guide
by integrating over the atom-beam profile. At a current of 3.0 A,
we guide up to $2\cdot10^6$ atoms/sec.

Figure 3 shows the atom-flux dependence on the track wire current.
For low currents ($<$0.7 A), the guiding potential should be
sufficient to confine the initial transverse-velocity
distribution, but it does not provide sufficient force to bend the
atoms around the curve; hence no flux is observed at the detector
(Fig. 3). Above the estimated centripetal track current threshold
of 0.7 A, atoms with low longitudinal velocity are guided. At
currents above 2.3 A, the flux saturates as we have sufficient
magnetic gradient to guide all longitudinal velocities around the
bend.

In our design, the magnetic-field minimum is 50 $\rm{\mu m}$ above
the substrate.  By lowering the position of the magnetic-field
minimum, and thus the guiding center of the track, to near or
below the substrate, we should be able to guide the atoms into the
glass. We do this by applying a bias field transverse to the track
and parallel to the substrate with a pair of Helmholtz coils.
Depending on the polarity of our Helmholtz coils, the minimum is
either raised or lowered. This variable bias field is set up at
the final 1-2 cm stretch of our guide, after the atoms have
negotiated the bend.  We observe that for one polarity of the
coils the guided flux changes by $\sim10 \%$, but as the bias
field polarity is reversed, the flux is completely eliminated
(Fig. 4). In the first case, the bias field lifts the field
minimum out of the tracks, but atoms still make it to the hot
wire, although they are shifted vertically. In the second case, we
lower the field minimum and atoms are guided into the substrate
surface. Once atoms touch the substrate, they stick and can no
longer be guided. As expected, larger track currents require
larger bias fields to push atoms into the substrate (Fig. 4). The
bias field required to entirely squelch the guided-atom flux is
directly determined by the track current. The three track currents
of 1.25, 2.0 and 3.0 A should correspond to squelching bias fields
of 21, 33, and 49 Gauss, respectively. Our bias-field measurement
shows that for the above track currents 21, 34, and 43 Gauss are
necessary to cut off the flux, which is in good agreement with the
estimates .

We measure the guided atoms' transverse velocity profile by
translating the hot wire to map out the spatial extent of the atom
beam as it diverges from the exit of the track (Fig. 5). We
calculate that the atoms' emergence from the confining fields of
the tracks is almost completely non-adiabatic---the transverse
kinetic energy of the emerging beam should thus be a faithful
reflection of the transverse kinetic energy in the guide. The
virial theorem tells that the mean potential energy in the linear
confining field should be twice the mean kinetic energy. The total
mean energy per transverse dimension of the atoms in the guide
should thus be three times the observed kinetic energy of the
emerging atoms.  At a track current of 2 A, we can determine that
the guided atoms have $3\times42\rm{\mu K}=126\rm{\mu K}$ total
mean energy per transverse dimension (Fig. 5). This is smaller
than the lowest point on the rim of the confining potential, which
is 1.1 mK.

In the discussion above we apply a bias field parallel to the
substrate and perpendicular to the track.  Similarly, we can apply
a transverse bias field in the vertical direction, perpendicular
to the substrate. In this configuration the guided atom beam is
moved close to one of the wires and the highest-energy component
of the beam is "skimmed off". Figure 5 shows a case in which the
mean energy in the corresponding direction has been reduced by a
factor of 3.1 using a 43 Gauss bias field.

Due to the non-adiabatic loading into our tracks, atoms entering
the center portion of the guide are more likely to be guided.  The
effective aperture through which atoms can enter the guide is
consequently smaller than the physical wire separation.  Comparing
the estimated source flux to the number of atoms detected after
the guide, we estimate that  $\sim 25\%$ of the $m_F=-1$ atoms
that hit the $100\times100$ $\rm{\mu m}$ opening are detected.

In summary, we have guided a beam of laser-cooled atoms between
two $100\times100$ $\rm{\mu m}$ wires.  We are able to bend the
atoms' trajectory around 15-cm-radius curves. The guiding-current
threshold agrees with our theoretical prediction to within our
uncertainty from the atoms' longitudinal velocity. We demonstrate
that moderate current densities give guiding potentials of several
milli Kelvin. No external cooling for the wires was necessary. Our
magnetic guide design has great promise for applications in atom
interferometers due to its versatility and simplicity.  We tested
the tracks with current pulses up to 8 A before substrate heating
became a problem.  This test indicates that we can reduce the
radius of curvature by a factor of 4 to achieve a 3-4 cm track
radius. In future experiments with larger currents, we hope to
guide the atoms around a full 360 degree bend, which is an
important step for a possible future Sagnac interferometer using
magnetic-field confinement. Photolithographic technology will
provide a reproducible method for producing an atom beamsplitter
in future atom-interferometer applications.

The authors would like to thank Carl Wieman and Eric Abraham for
helpful discussions.  This work was made possible by funding from
the Office of Naval Research (Grant No. N00014-94-1-0375) and the
National Science Foundation (Grant No. Phy-95-12150).

\pagebreak

\epsfxsize=6 truein \epsfbox{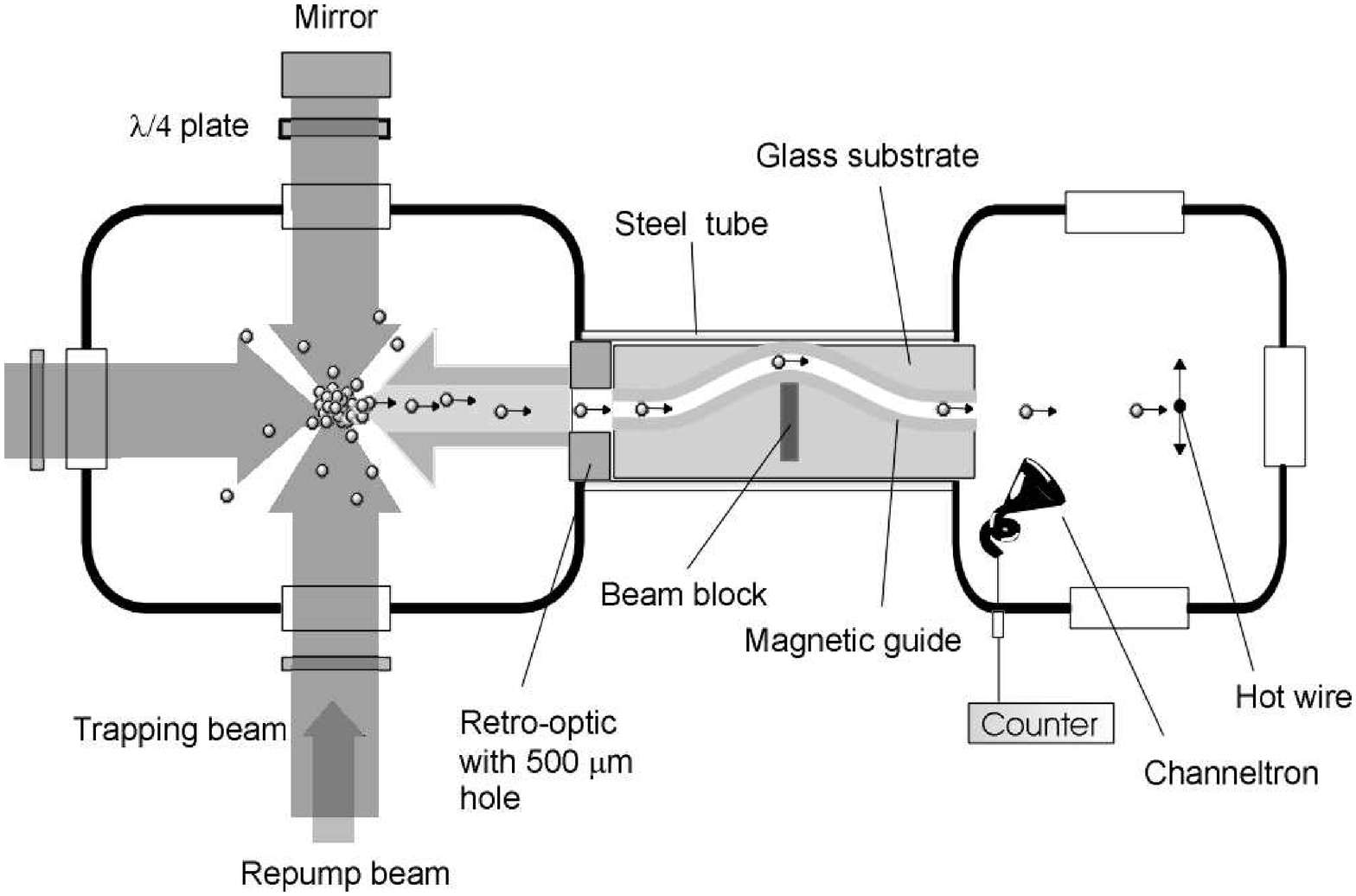}
\begin{figure} \caption{Schematic of experimental setup.  A beam
of laser-cooled atoms generated in the source chamber (left)
travels toward the magnetic guide between the source and detection
chamber. Atoms enter the tracks and are guided around a barrier.
After the guided atoms exit the tracks, they are ionized with a
hot wire and ions are counted with a channeltron. \label{fig1}}
\end{figure}

\epsfxsize=5 truein \epsfbox{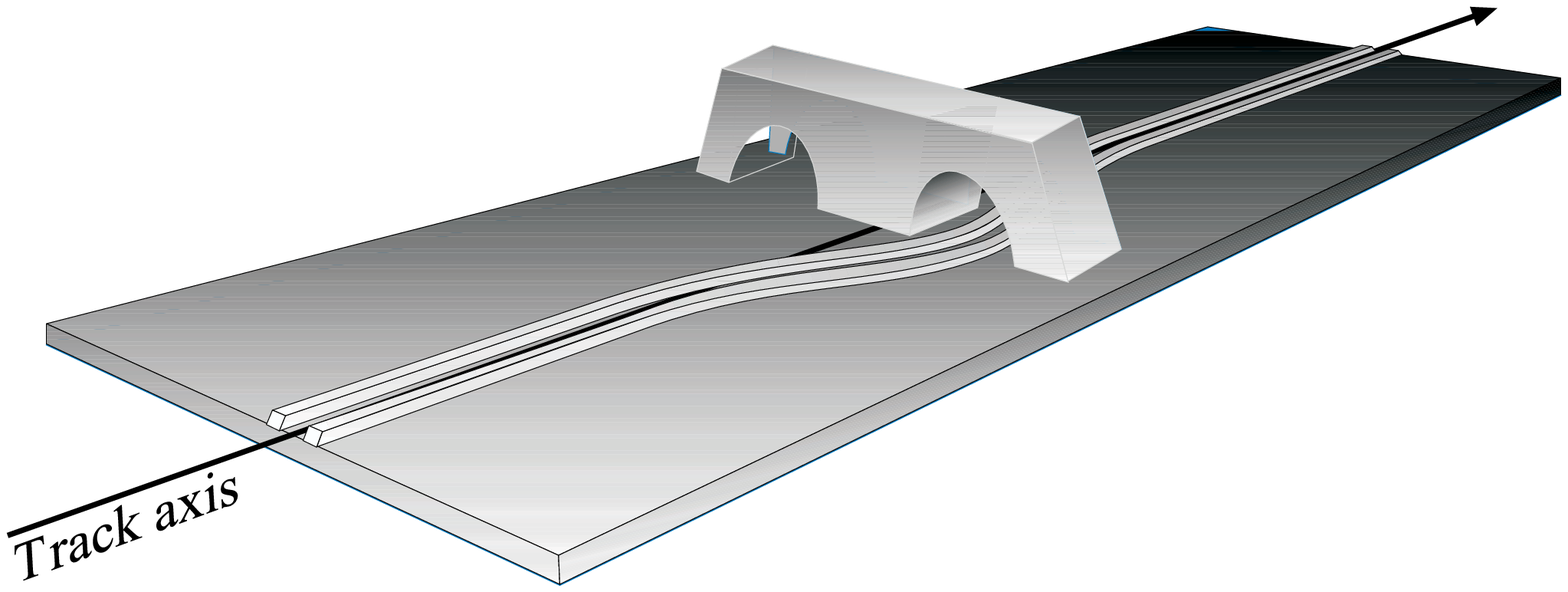}
\begin{figure} \caption{Detail of magnetic guide.  Atoms are
guided over a 10-cm distance around three curves each with a 15-cm
radius of curvature. A beam block in the middle of the guide
blocks the center of the LVIS beam. The transverse scale of the
track has been exaggerated for clarity. \label{fig2}}
\end{figure}

\pagebreak

\epsfxsize=4 truein \epsfbox{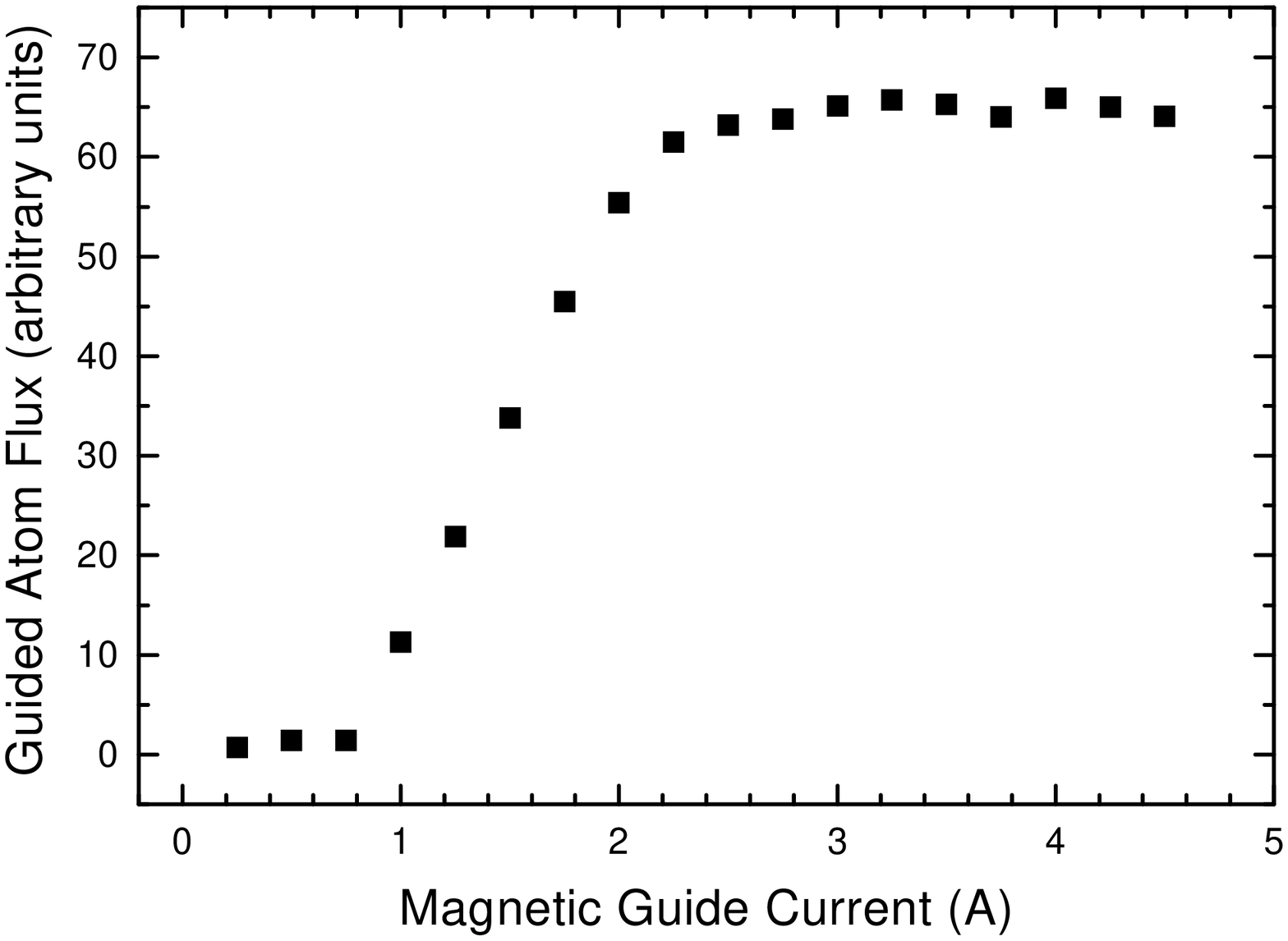}
\begin{figure} \caption{Flux versus current.  As we increase the
current in the guiding wires the magnetic field and its gradient
between the wires increases and atoms are guided.  For low
currents ($<0.7$ A), the longitudinal velocity of the atoms is too
large to be bent around the curve.  We increase the current, and
the guiding flux increases until saturating ($>2.3$ A), when all
atoms between the wires are guided.  Each data point represents
the average of 50 current pulses; the statistical errors are
typically smaller than the plotted symbols. \label{fig3}}
\end{figure}

\epsfxsize=4 truein \epsfbox{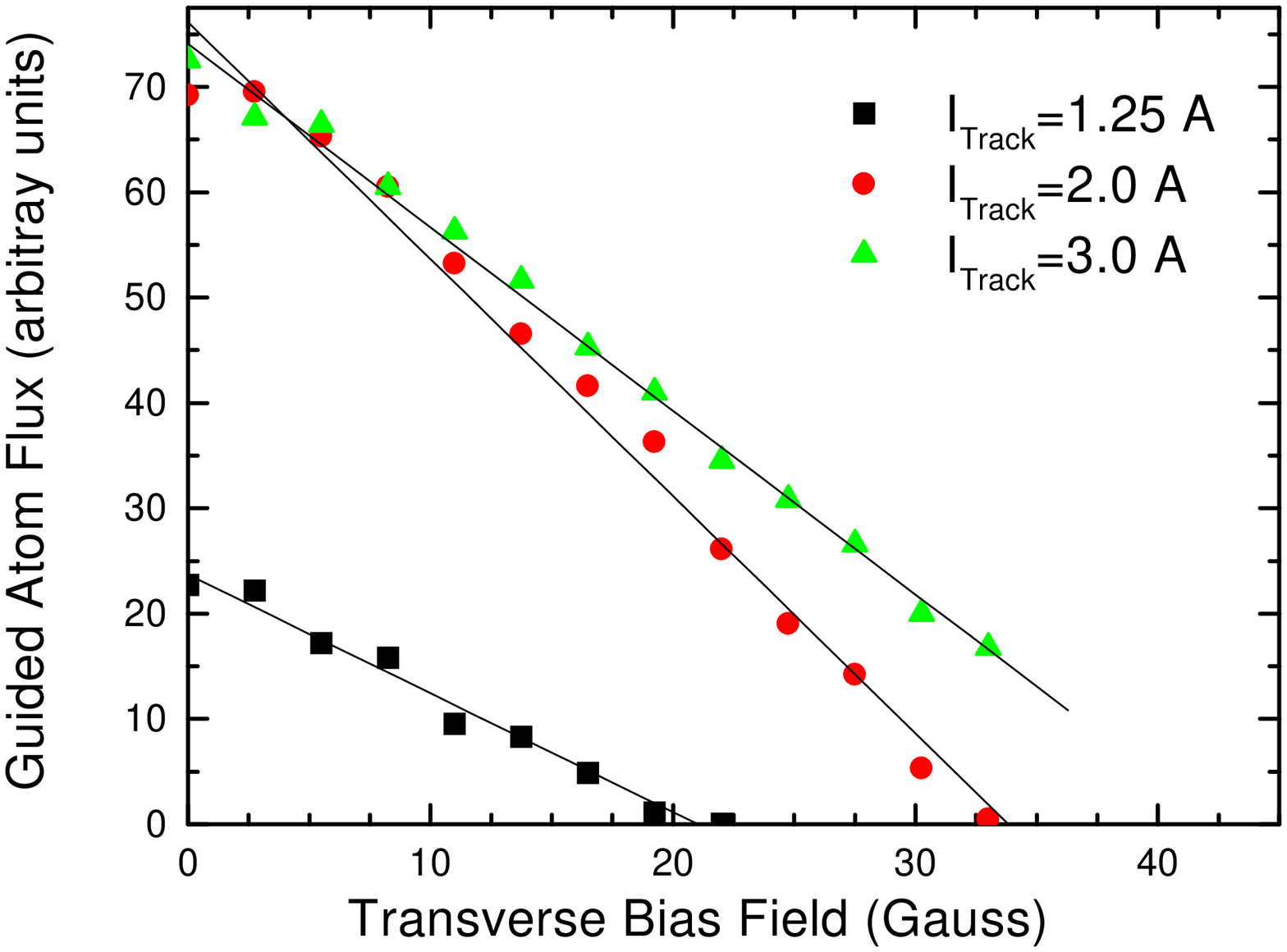}
\begin{figure} \caption{Guided flux versus bias field parallel to substrate.  When a bias
field is applied parallel to the guide, the minimum of magnetic
field is lowered toward the substrate.  We lower the
magnetic-field minimum until the atoms run into the track
substrate and the flux is extinguished. The solid lines represent
a linear fit to the data. Larger track currents require larger
bias fields to cut off the flux. \label{fig4}}
\end{figure}

\pagebreak

\epsfxsize=4 truein \epsfbox{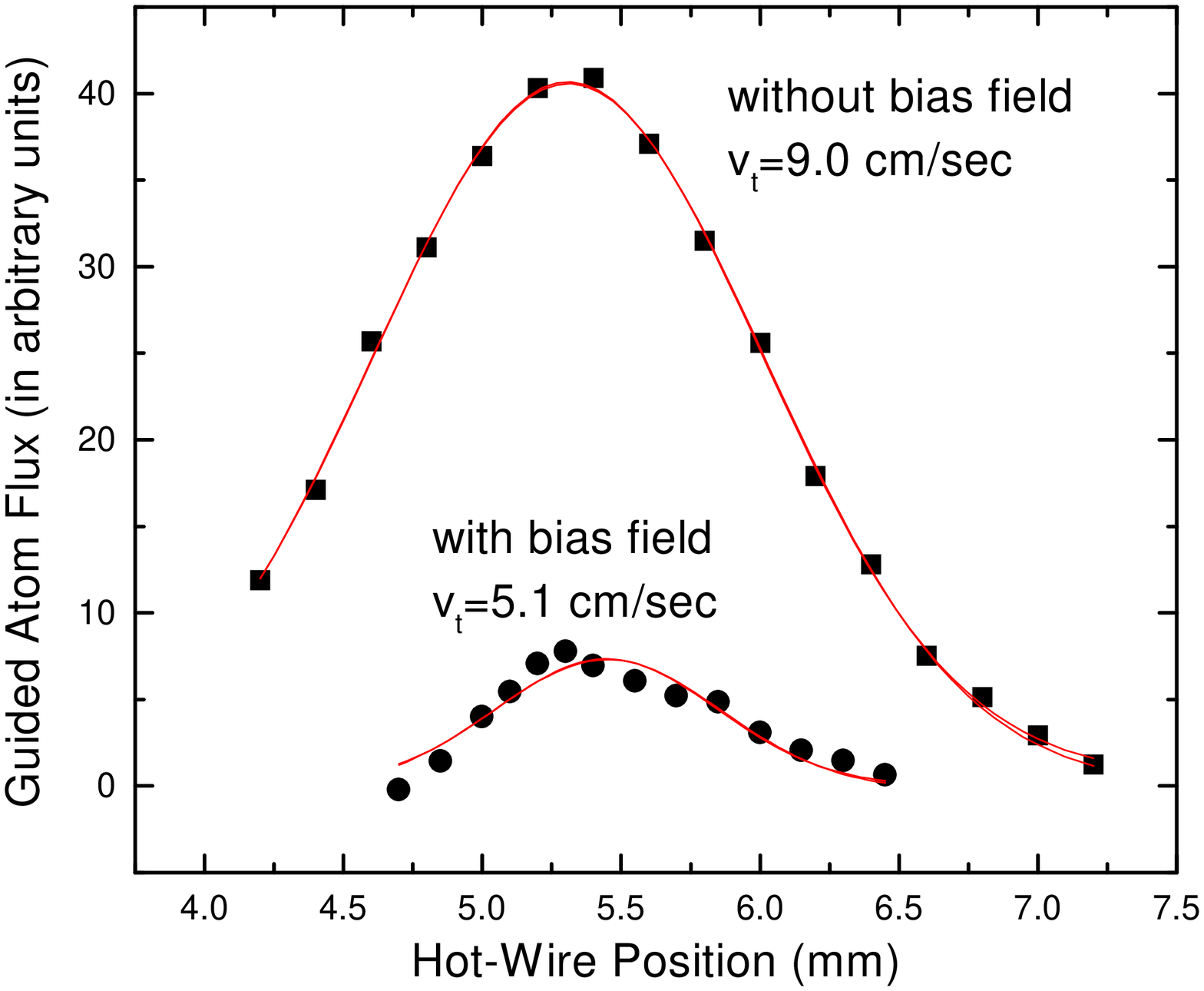}
\begin{figure}
\caption{Transverse-velocity profile. We translate the hot wire
across the guided-atom beam and measure the transverse-velocity
distribution.  A 43 Gauss bias field applied in the vertical
direction moves the minimum of magnetic field toward one of the
wires, and the highest-transverse-energy component is "skimmed
off". We use the width of the Gaussian fit to determine the RMS
transverse velocity $v_t$. \label{fig5}}
\end{figure}


\begin{references}
\bibitem{magnetic guide} As this manuscript was being completed,
we learned that the Harvard atom-optics group has seen a signal
for flux through patterned wire guides.  M. Prentiss, private
communication.
\bibitem{Metrigraphics}  Metrigraphics is a division of Dynamics Research
Corporation, Wilmington, MA. Metrigraphics is a trade name used
here for identification purposes only and does not constitute an
endorsement by the authors or their institutions.
\end{references}
\end{document}